\documentclass[prl,twocolumn,superscriptaddress,showpacs]{revtex4-1}
\usepackage{bm}
\usepackage{amsmath}
\usepackage{amssymb}
\usepackage{amsxtra}
\usepackage{amscd}
\usepackage{amsthm}
\usepackage{amsfonts}
\usepackage{eucal}
\usepackage{latexsym,amsthm}
\usepackage{color}
\usepackage{graphicx}

\usepackage{natbib}

\usepackage{setspace}

\newcommand{\nc}{\newcommand}
\nc{\on}{\operatorname}
\nc{\wt}{\widetilde}
\nc{\Wick}{{\mathbb :}}
\nc{\R}{{\mathbb R}}

\newcommand{\beq}{\begin{equation}}
\newcommand{\eeq}{\end{equation}}
\newcommand{\bmul}{\begin{multline}}
\newcommand{\emul}{{\end{multline}}}
\newcommand\beqa{\begin{eqnarray}}
\newcommand\eeqa{\end{eqnarray}}
\newcommand\bea{\begin{array}}
\newcommand\eea{\end{array}}
\newcommand\ba{\begin{array}}
\newcommand\ea{\end{array}}
\newcommand{\nn}{\nonumber}

\newcommand{\neqa}{\nonumber\end{eqnarray}}

\newcommand{\Eq}[1]{Eq.(\ref{#1})}

\newcommand{\ur}[1]{(\ref{#1})}

\renewcommand{\d}{\partial}

\nc{\CH}{{\mathcal H}}
\nc{\Db}{{\bar D}}
\nc\comment[1]{}

\nc{\CM}{{\mathcal M}}
\nc{\CN}{{\mathcal N}}

\newcommand{\re}{\relax{\rm I\kern-.18em R}}

\renewcommand{\Im}{{\mathrm {Im}}}
\renewcommand{\Re}{{\mathrm {Re}}}
\nc{\meV}{{\mathrm{\,meV}}}
\nc{\cG}{{\mathcal G}}

\renewcommand{\)}{\right)}
\renewcommand{\(}{\left(}
\renewcommand{\bar}{\overline} 

\nc{\al}{{\alpha}}

\def\eps{{\epsilon}}
\begin{document}

\title{Effect of paramagnetic fluctuations on a Fermi surface topological transition in two dimensions}
\author{
Sergey Slizovskiy}
\email{on leave from PNPI; S.Slizovskiy@lboro.ac.uk} 
\author{Joseph J. Betouras}
\email{J.Betouras@lboro.ac.uk} 
\affiliation{Department of Physics, Loughborough University, Loughborough LE11 3TU, UK}
\author{Sam T. Carr} \affiliation{Institut f\"ur Theorie der
  Kondensierten Materie and DFG Center for Functional Nanostructures,
  Karlsruher Institut f\"ur Technologie, 76128 Karlsruhe, Germany}
   \affiliation{SEPnet and Hubbard Theory Consortium, School of Physical Sciences, University of Kent, Canterbury CT2 7NH, UK}
\author{Jorge Quintanilla} \affiliation{SEPnet and Hubbard Theory Consortium, School of Physical Sciences, University of Kent, Canterbury CT2 7NH, UK}

\begin{abstract}
We study the Fermi surface topological transition of the pocket-opening type in a two dimensional Fermi liquid.  We find that the paramagnetic fluctuations in an interacting Fermi liquid typically drive the transition first order at zero temperature. 
We first gain insight from a calculation using second order perturbation theory in the self-energy. This is valid for weak interaction and far from instabilities. We then extend the results to stronger interaction, using the self-consistent fluctuation approximation.  Experimental signatures are given in the light of these results.
\end{abstract}
\maketitle
\comment{$}
{\it Introduction.}---Fermi surface reconstruction has been a subject of fundamental interest in the theory of metals for a long time.  In non-interacting models, the Lifshitz transition \cite{Lifshitz} is a well known example of a Fermi surface topological transition (FSTT), where the Fermi surface (FS) changes its topology when some external parameter (for example, pressure) is varied.  Such FSTTs may change the number of FSs (for example in a pocket-opening transition), or the nature of a single FS (for example in a neck-closing transition).  An important characteristic of these transitions however is that they occur \textit{without} symmetry breaking.  

Rather than consider an external agent, one may further ask if such FS deformations can occur as a function of some interaction strength.  In this case, one may have symmetry breaking transitions (Pomeranchuk instabilities \cite{Pomeranchuk}), as well as non-symmetry breaking deformations, predicted in early works \cite{Kohn, Nozieres}.  It has since been understood that Pomeranchuk and interaction-driven topological transitions can be put on the same footing \cite{Quintanilla}.  It is then natural to look for a unified theory of the non-interacting Lifshitz FSTTs tuned by an external agent and the FS distortions induced by interactions.

The interest regarding FSTTs in solids has surged recently, with a plethora of works
both theoretical \cite{Quintanilla,Carr, Katsnelson, Hackl,
Chen, Yamaji, LeeStrackSachdev,ChubukovMorr} and experimental \cite{Hiroi, Liu}, in contexts ranging from nematic phases in cold atoms \cite{Carr} to AFM fluctuations in cuprates \cite{ChubukovMorr}.
In this work, we address the general effects of paramagnetic (PM) fluctuations on a FSTT, specifically asking what is the order of the transition and what are the experimental signatures?
The answers to these questions are crucial for the understanding of a number of materials, but this study is especially motivated by the layered material, Na$_x$CoO$_2$, which is known to both contain strong PM fluctuations, and have a band structure that admits a FSTT as a function of doping \cite{Hiroi}.

\begin{figure}[t] 
\begin{center}
\includegraphics[scale=0.35]{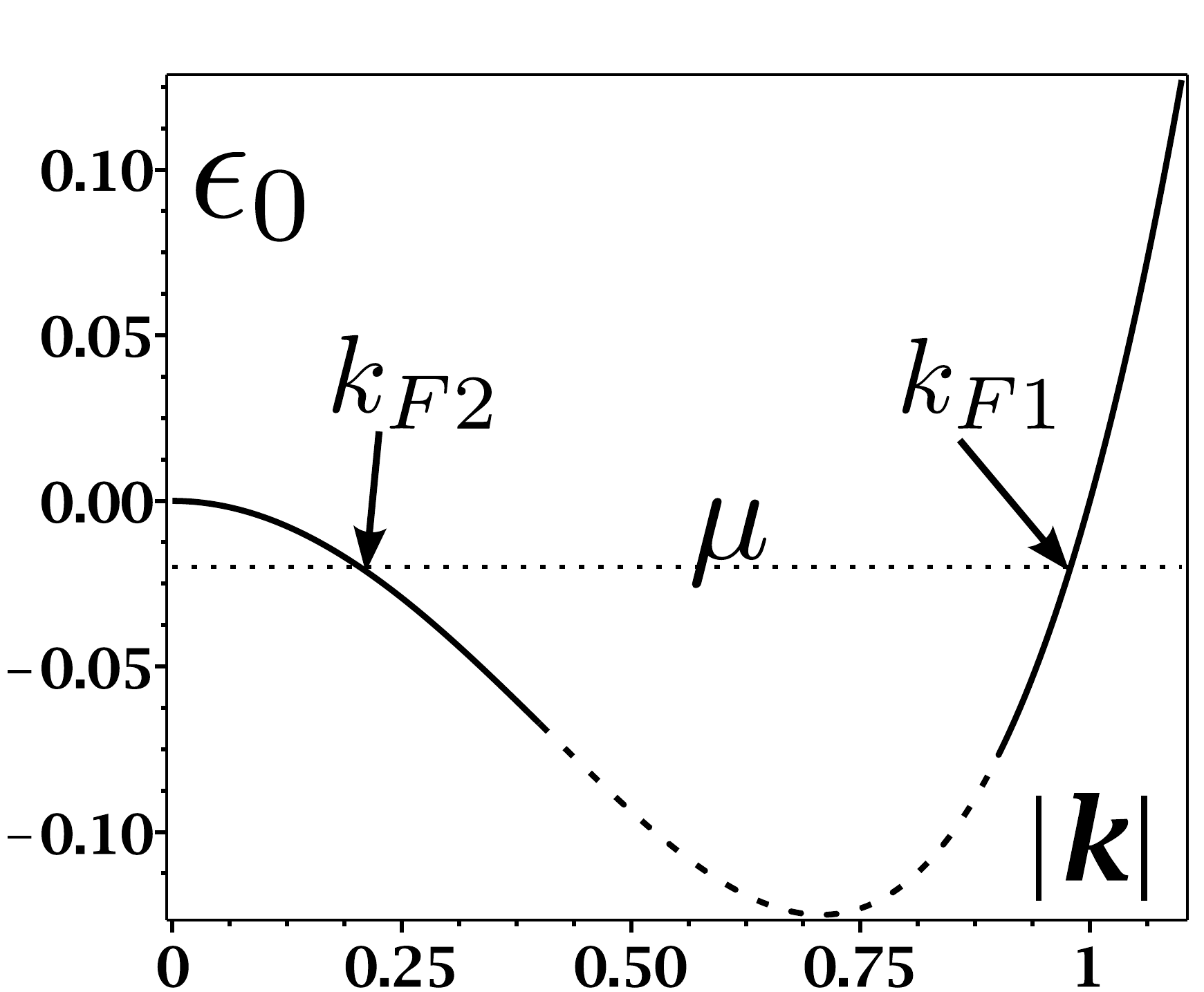}
\end{center}
\vspace{-0.5 cm}
\caption{\label{fig:dispersion} 
Schematic representation of the bare dispersion\label{fig:two solutions} }
\vspace{-0.6 cm}
\end{figure}

The fundamental physical feature of the problem is a non-interacting dispersion relation which leads to a large (and for simplicity, circular) FS in two
dimensions (2D), but with the chemical potential close to the energy dispersion relation at the centre of the band (take $\mu \approx 0 $) as shown in Fig.~\ref{fig:dispersion} so that in the non-interacting case, a small pocket may appear as a function of doping. In the following, we will calculate the effect of short-range interactions on this FSTT.
We seek to characterize this transition, and show its signatures in various physical quantities.  

The system, we consider, is a Fermi liquid (FL) with short-ranged (Hubbard type for simplicity) interactions. The Hamiltonian becomes then: 
$H = \sum_{\sigma=\uparrow, \downarrow} \int {\rm d}^2 k/(2\pi)^2 [ \eps_0({\bf{k}}) c_{{\bf{k}},\sigma}^\dag c_{{\bf{k}},\sigma} 
+  U n_{\uparrow}({\bf{k}}) n_{\downarrow}(-\bf{k})]$, 
where $\epsilon_0(\bf{k})$ is the kinetic energy, $c_{\bf{k},\sigma}^\dagger$ and $c_{\bf{k},\sigma}$ the creation and annihilation operators in momentum space, and $n_{\sigma}({\bf{q}})= \int \frac{d^2 k }{(2\pi)^2} c_{\bf{k}+\bf{q},\sigma}^\dag c_{\bf{k},\sigma}$ the density of electrons with spin $\sigma=\uparrow,\downarrow$. As is typical in field theories in 2D, we assume an essential cut-off momentum for the interaction $\Lambda$ \cite{AGD}. This cut-off can be physically related to the inverse interparticle distance or the inverse screening length of a more realistic short-ranged potential.

The dispersion close to the centre of the FS can be approximated in dimensionless units by $\epsilon_2({\bf{k}}) = - \frac{k^2}{2}$ while for the large circular FS
the dispersion, including the curvature term, is $\epsilon_1({\bf{k}}) = v_{F1} (|k|-1) + \frac{(k-1)^2}{2 m_1}$ (we use units where  $k_{F1}=1$ at $\mu=0$).
Without interactions, the pocket appears continuously by changing the doping (chemical potential $\mu$), the magnitude of its Fermi vector is denoted by $k_{F2}$.
With interactions however, there is competition between the kinetic energy $E_{kin} \sim k_{F2}^2$ and the
interaction contribution to the self-energy; this competition is crucial in determining whether one may add a particle to the small FS or not.

It is instructive to first study perturbation theory only to second order (SOPT) in the self-energy, far from any symmetry breaking instabilities and for very weak interaction.  We will then use the developed insight, to concentrate on the region of paramagnons near a ferromagnetic (FM) instability, which is the central part of this work.

{\it Second order perturbation theory.}---
To first order in perturbation theory, we obtain the Hartree self energy $\Sigma(k_F,\Omega=0)=Un/2$ (the Fock term is zero), which can be absorbed in $\mu$.
The first non-trivial order is therefore the SOPT:
\beq \label{SigmaU2}
\Sigma(k_F,\omega=0) = U^2 \int \frac{d^2 q d\omega'}{(2 \pi)^3} G({\bf{k_F}}+{\bf{q}}, i \omega') \chi_0(-{\bf{q}},-i \omega') 
\eeq
\noindent where $\chi_0$ is the susceptibility of free fermions.

Assuming the onset of a small pocket, $\chi_0$ contains contributions from both the large and small FSs.  It will turns out however that the important physics we want to reveal occurs at the small FS -- so we initially ignore the large FS and concentrate only on the pocket.  The contribution to the susceptibility from the small FS and for $q \gg k_{F2}$ reads
\beq
 \chi_{0}({\bf{q}},i \omega) \approx \frac{k_{F2}^2}{\pi (q^2 + 4 \frac{\omega^2}{q^2})},
\eeq
which leads with logarithmic accuracy to \cite{comment3}
\beq \label{SelfEnergySmallU}
\Sigma(k_{F2},i \omega=0) \approx \frac{U^2}{8 \pi^2} k_{F2}^2 \ln\frac{\Lambda}{k_{F2}} + const.
\eeq
This strong logarithmic dependence on $k_{F2}$ is crucial in the determination of the size of the pocket. 
 For a given chemical potential $\mu$ (we assume the smooth Hartree term is already absorbed) this is given by the solution of the energy balance equation
\beq \label{simple-balance}
\mu= \eps(k_{F2}) + \Sigma(k_{F2},\omega=0) = \frac{k_{F2}^2}{2} \left( \frac{U^2}{4 \pi^2}  \ln\frac{\Lambda}{k_{F2}} -1 \right).
\eeq  

Consider $\mu>0$, so that the non-interacting model has all small-momentum states filled and there is no pocket. 
 In the presence of interactions however, the effective dispersion for $k_{F2}$
bends slightly up, with a maximum $\mu_{max}=\frac{U^2}{16 \pi^2} \Lambda^2 e^{-\frac{8 \pi^2}{U^2} - 1}$  
reached at  $k_{F2} = \Lambda e^{-\frac{4 \pi^2}{U^2}-\frac{1}{2}}$.
In addition to the trivial solution without a pocket, there are then two non-trivial solutions of \Eq{mu equation} for $k_{F2}$ in the interval $\mu\in[0,\mu_{max}]$, as seen
in Fig.\ref{fig:two solutions}b.  Solving for $k_{F2}$ yields
\beqa \label{kF2-solution}
k_{F2}^{(1);(2)}(\mu) = \frac{4 \pi  \sqrt{\mu }}{U \sqrt{-W_{0;-1}\left(-\frac{16 \pi ^2 \mu  e^{\frac{8 \pi
^2}{U^2}}}{\Lambda^2 U^2}\right)}},
\eeqa     
where $W_i(z)$ is the product logarithm function \cite{comment2}.
  
\begin{figure}[t] 
\begin{center}
\includegraphics[scale=0.23]{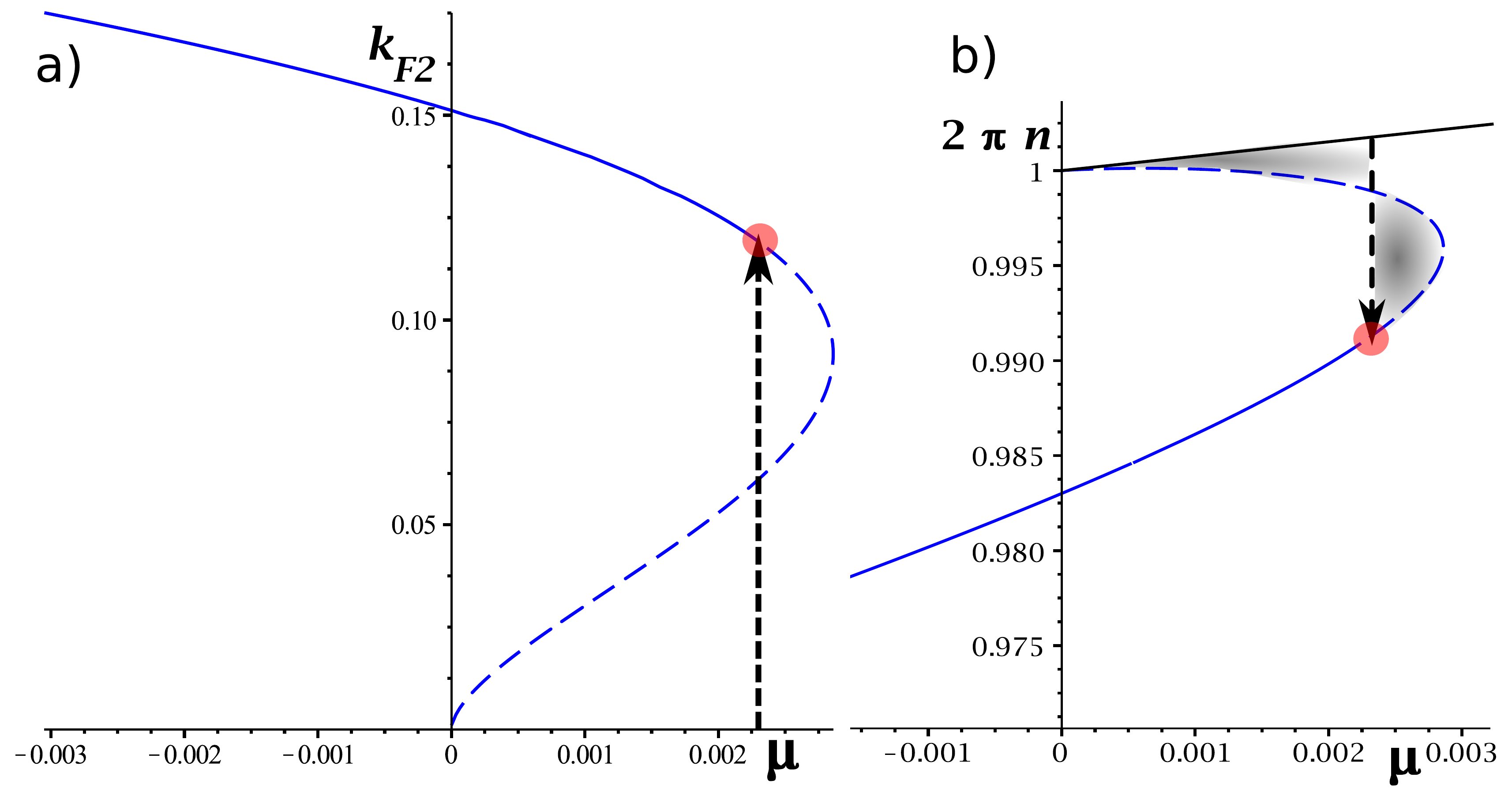}
\end{center}
\vspace{-0.7 cm}
\caption{\label{fig:Lifshitzbig} 
Plots of 
a) Pocket size $k_{F2}(\mu)$, 
b) Electron density $n(\mu)$,  
for parameters $U=4$ (effective $U_{\rm{eff}}=6.3$), $v_{F1}=2$, $m_1=1$, $\Lambda=1$ (other parameters give qualitatively similar results).
For these values the FSTT happens at $\mu=0.0023$ and electron density jumps by roughly $1\% $.
The dashed arrow indicates the phase transition, which happens when the two shaded domains have equal area, reflecting a Maxwell construction. 
}
\end{figure}

Which of the three solutions for $k_{F2}$ is stable is determined by the grand canonical potential $\Omega$,
which we find by integrating  ${\rm d}\Omega = - n \, {\rm d}\mu$ starting from the point $k_{F2}=0$ where
the phases merge. 
The result is plotted in Fig.~\ref{fig:Lifshitzbig}: the trivial phase with $\mu<\mu_{crit}$ is unstable and a first order
phase transition to the solution with a larger pocket happens (indicated by arrow). 
This
happens because at small $k_{F2}$ the logarithm in the $U^2$ term outweighs the free kinetic term.  From the expression for $\Omega$ it follows that the position of $\mu_{crit}$ divides equally the shaded area; a case of Maxwell construction.

Although the above picture of FS reconstruction as a function of interaction $U$ as in Refs.~\cite{Kohn, Nozieres} is intuitively appealing, in a typical experimental situation one rarely has any strong control over $U$.  We therefore imagine returning to the Lifshitz setup, where some external parameter is varied, extending these original ideas to non-zero interaction.  With the concrete example of Na$_x$CoO$_2$ in mind, we examine what happens as a function of doping.  To make this picture consistent however, we must first reinstate the large FS.

%
%

It may be intuitively expected that the self-energy of the large FS has no essential dependence on the size of the pocket; however to check
this we evaluate this contribution 
\beq
\Sigma(k_{F1},\omega=0)  \approx \frac{U^2}{8 \pi^2} \, k_{F2}^2 \frac{1-v_{F1} m_1}{m_1 v_{F1}^2}.
\eeq 
 
\noindent which has no logarithmic enhancement. At the level of SOPT, the main role of the large FS is therefore to act as a particle reservoir, with the electron density given by Luttinger's theorem: $n = \frac{1}{2 \pi} (k_{F1}^2 - k_{F2}^2)$. 

We assume that all the non-divergent terms containing $k_{F2}^2$ are effectively included in $\Lambda$ and the small self-energy contributions to the large FS are already included in the bare dispersion parameters.  We also assume that the effective chemical potential $\mu$ already includes the Hartree term, as this can play no role at constant density. Then, the chemical balance equation reads:
\beq \label{mu equation}
 \mu= \eps(k_{F2}) + \Sigma(k_{F2},\omega=0) = v_{F1} (k_{F1}-1 ).
\eeq
\noindent 
This has a solution $k_{F1} = \frac{\mu}{v_{F1}}+1$, while $k_{F2}$ is given from Eq.~\eqref{kF2-solution} and the discussion above.  
However, one may now view the pocket-appearing transition as a function of density (parameterized by $\mu$).
Without interactions, the pocket smoothly appears for $\mu<0$. In the presence of interactions however, 
the point of a non-interacting FSTT ($k_{F2}=0 ,\ \mu = 0$) is unstable and
when $\mu$ is slightly above zero, there is a first-order phase transition to the branch with larger $k_{F2}$. We see however from Eq.~\eqref{kF2-solution} that for small $U$, the jump of $k_{F2}$ at the phase transition
is exponentially small.
We also note that the two phases have a different electron density which decreases abruptly by $\frac{k_{F2\ min}^2}{2 \pi}$ at the FSTT
\cite{comment4}.

To summarize so far, interactions drive the pocket-vanishing FSTT first order in SOPT, with an exponentially small jump of $k_{F2}$ for small $U$. Going to higher orders in $U$ in general, where a small FS in 2D may have further non-analyticities due to other fluctuations (see e.g. Ref.~\cite{Bloom, Belitz}), is beyond the scope of the this study.  However, by concentrating on the region of large PM fluctuations, we will now show that this jump is enhanced for larger $U$.

{\it Moderate U.}---By increasing the strength of the interaction, approaching but remaining below the Stoner instability, we enter the regime of paramagnons.
The summation of ladder and ring diagrams \cite{Doniach} gives the "effective paramagnon" interaction:
\beq \label{V}
V({\bf{q}},i \omega) = \frac{\chi_0({\bf{q}},i \omega)}{1-U^2
\chi_0^2({\bf{q}},i\omega)} + \frac{U \chi_0^2({\bf{q}},i \omega)}{1-U \chi_0({\bf{q}},i \omega)}
\eeq 
The self-energy in the paramagnon approximation in the low temperature limit then reads: 
\beqa \label{Sigma real}
&\Sigma({\bf{k}},i \Omega_n) = 
\\ &  U^2 \int_{-\infty}^\infty d \omega \int \frac{d^2q}{(2 \pi)^3} G({\bf{k}}+{\bf{q}},  i \Omega_n + i \omega) V({\bf{q}},i \omega).
\nn
\eeqa
Further diagrams giving the vertex corrections turn out to cancel with those of the qp weight $Z$
\cite{Hertz, Hansch}.

\begin{figure}[t] 
\begin{center}
\includegraphics[scale=0.35]{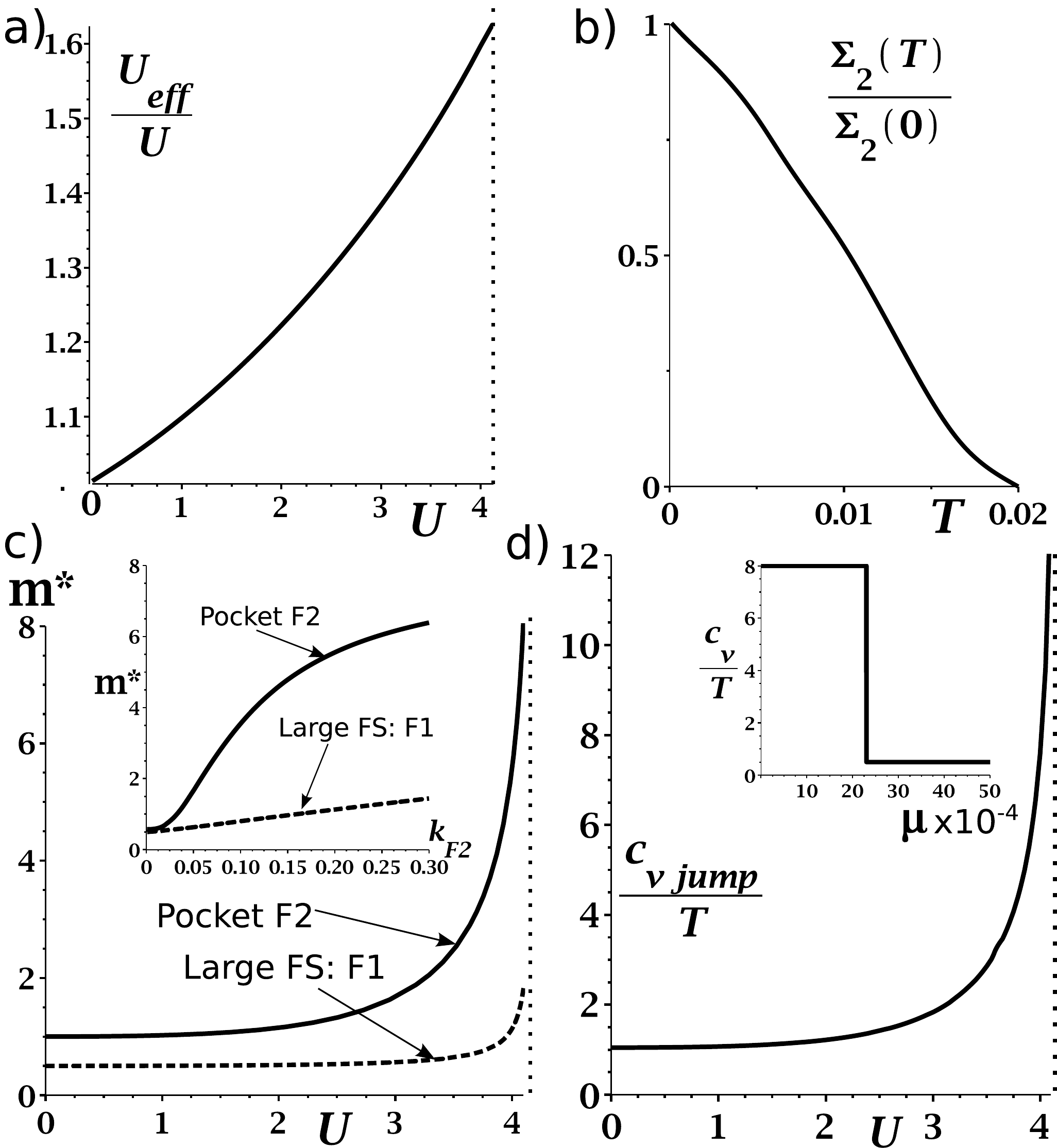}
\end{center}
\vspace{-0.35 cm}
\caption{\label{fig:Ueff} 
a) Plot of $U_\mathrm{eff}/U$ as a function of $U$, defined through \Eq{kF2Fit}  b) Temperature dependence of
$\Sigma(k_{F2})$ for fixed $k_{F2}$ c) Effective mass at both FSs as a function of $U$ ($k_{F2}=0.2$)  and as a function of $k_{F2}$ for
$U=4$ in the inset (solid and dashed lines refer to small and large FS respectively)
d) Jump in T-linear coefficient in $c_v$ due to FSTT for $v_{F1} = 2$.
}
\end{figure}

As before, we express the bare susceptibility as a sum of two Lindhard functions, coming from the two FSs.
 Evaluating numerically the integral \Eq{Sigma real} for the real part of the self-energy at the small FS, we find that it can be well fitted by the function
\beq \label{kF2Fit}
\Sigma(k_{F2}) \approx \frac{U_\mathrm{eff}^2}{8 \pi^2} \left[k_{F2}^2 \(\ln\frac{\Lambda}{k_{F2}} + a_1\) + \frac{b_1}{v_{F1}} k_{F2}^2 + c_1
\right]. 
\eeq
Here, $a_1$ slowly increases with $U$ and is close to zero for large $U$,  and $b_1 \approx -0.8$. 
The first two terms come from the contribution of the susceptibility from the small FS, while the $b_1$ term from the large FS
susceptibility.  Aside from the small analytic corrections, $a_1,b_1$, the overall form of the self-energy is identical to Eq.\ur{SelfEnergySmallU}, with an effective interaction strength, $U_\mathrm{eff}$.  
For small $U$, $U_\mathrm{eff} = U$, but when $U$ approaches the Stoner instability $U_\mathrm{Stoner} = 2 \pi/(1+1/v_{F1})$, the effective interaction strength greatly increases.  This is plotted in Fig.~\ref{fig:Ueff}a, and 
can be understood as the effect of the enhancement of the susceptibility due to PM fluctuations.  
The large FS now plays a role beyond that of a reservoir: it drives the system closer to the Stoner instability, thus enhancing the PM fluctuations and consequently $U_\mathrm{eff}$.
The physics of the first-order Lifshitz transition described previously by SOPT is still valid, with the simple replacement of $U\rightarrow U_\mathrm{eff}$.

It is worth mentioning that while a similar effect of the interactions driving the FSTT first order has previously been discussed for neck-opening
transitions in Ref. \onlinecite{Carr}, the physical processes in the present case are quite different.  In the former, the
logarithmic divergence of the single-particle density of states (van-Hove singularity) led to a first order transition
technically originating from the Fock term.  In the present study, the large paramagnetic fluctuations lead to the effect.



{\it Properties.}---Having shown that the Lifshitz transition is driven 1st order by interactions, we now turn to the physical consequences of this calculation.  We begin by addressing the question: is the system with the small FS (and large self-energy) a good FL?
To answer this, we consider the frequency and momentum dependence of the retarded self-energy which reads \cite{ChubukovMaslov2003}:
\beqa \label{SigmaIm1}
\nonumber
\Im \Sigma^R({\bf{k}},\omega) &\approx&  U^2 \int \frac{d^2q}{(2 \pi)^2} \left[\theta(\eps({\bf{q}})) -
\theta(\eps({\bf{q}})-\omega)\right]  \\
&\times& \Im V^R({\bf{q}}-{\bf{k}},\eps({\bf{q}})-\omega)
\eeqa

First in SOPT, for the small FS and small momentum transfer (forward scattering) $\Im V^R(q,\omega) =\Im \chi_0^R(q,\omega)$ and it behaves as:
\beq
 \Im \chi_0^R({\bf{q}},\omega) = - \frac{1}{2 \pi m} \frac{\omega}{\sqrt{(k_{F2} q)^2 - \omega^2}} \theta(k_{F2} q - |\omega|), 
\eeq
The computation reveals a logarithmic term coming from the region of forward scattering. 
A similar one comes from the backscattering region $q \approx 2 k_{F2}$. Therefore, when only the small pocket is considered, 
with logarithmic accuracy  \cite{ChubukovMaslov2003}:
\beq \label{SigmaIm2}
\Im \Sigma(k_{F2},\omega) \approx \frac{U^2}{8 \pi^3 k_{F2}^2} \omega^2 \log \frac{k_{F2}^2}{|\omega|} 
\eeq

In the region of paramagnetic fluctuations, the imaginary part of
the self-energy behaves as in SOPT with $U$ replaced by $U_{eff}$, while the real part reads:
\beqa
 \Re \Sigma(k_{F2}, \omega>0) =
 \left\{\ba{cc} (1-Z^{-1}) \omega &, \omega \lesssim k_{F2}^2  \\
                    const.  & , \omega \gtrsim k_{F2}^2 \ea \right.
\eeqa 
The coefficients are $(-Z^{-1}+1) \approx - \frac{1}{8 \pi^2} U^2  V(0,0)$, with $V$ given by Eq.~\ref{V}.

These results show that in the case of small-$U$, where the pocket is exponentially small, it is, formally, a
good Fermi-liquid  with large quasiparticle weight $Z$. 
In addition, $Z$ dictates an evolution from good
Fermi-liquid at small $U$ to a bad Fermi-liquid close to Stoner instability. 
For example, for $U=4$, $v_{F1} = 2$ (as used in Fig.~\ref{fig:Lifshitzbig}) we obtain $Z^{-1} = 16.5$.
When $U$ increases, the effective mass
\beqa
m^* \equiv k_F /|v_F^*| \\
v_F^* = Z_{k_F} \(v_F + \frac{\d}{\d k} \Re \Sigma (k,0)\) 
\eeqa
diverges as a power-law close to Stoner transition. This can be clearly seen in measurements of the heat capacity, as discussed below, while the small $Z$ in this case leads to an almost smooth spectral function. This makes the pocket almost invisible in ARPES experiments. 

The effect of temperature $T$ is summarized in Fig.~\ref{fig:Ueff}b, where it is evident that the self-energy $\Sigma_2$ is
affected by $T$ and at reasonably high temperatures we can expect a termination of the effect. An accurate estimate of this temperature
will be given for realistic parameters elsewhere.

The heat capacity at low temperatures can be computed using the FL formula, which for 2D is:
\beq
c_v = \frac{\pi^2}{3} N^*(0) k_B^2 T = \frac{\pi T}{3} (m^*_1 + m^*_2) 
\eeq
where $N^*(0) = k_F/(\pi |v_F^*|)$ is the density of states at the Fermi level. For the non-interacting system when the pocket opens continuously, this leads to a jump in the coefficient of the term 
proportional to $T$ in $c_v$ at the FSTT. Also, $c_v$ depends only on the pocket's appearance and not on its size,
indicative of the 2D nature of the pocket.  In the presence of interactions, the PM fluctuations make $m^*$ of the
pocket dependent on its size for small $k_{F2}$. This dependence weakens for larger $k_{F2}$ (as seen in
Fig.~\ref{fig:Ueff}c), making the doping dependence of $c_v$ remarkably similar to the 2D non-interacting case but with the magnitude of
the jump at the FSTT dependent on the interaction strength $U$, as seen in Fig. \ref{fig:Ueff}d \cite{comment5}. 

However, for a real layered material such as Na$_x$CoO$_2$ which has a non-zero inter-layer hopping, the situation is different.  In this case, the non-interacting Lifshitz transition does not show a jump in $c_v$ as the pocket smoothly opens, instead it exhibits a square root singularity.  On the other hand, so long as the minimum $k_{F2}$ at the transition is larger than the inter-plane hopping, the theory developed above is unchanged. The predicted jump in $c_v$ was seen experimentally \cite{Hiroi}.  Thus, we believe that the present work provides the essential physics behind the FSTT in Na$_x$CoO$_2$.

Naturally, $c_v$ diverges at the FM transition due to the divergence of $m^*$ for both FSs, shown in Fig.\ref{fig:Ueff}c. 
As $U$ increases further, beyond the Stoner instability, the FSTT and the transition to a FM occur together; in this case the FSTT is driven first order \cite{Yamaji} by the magnetic transition.
 
Summarizing, we have demonstrated that interactions may drive a pocket opening FSTT first order in a 2D FL, in the region of PM fluctuations. 

We thank Andrey Chubukov, Piers Coleman, Dima Efremov, Sriram Shastry and Mathias Vojta for many discussions and suggestions. In particular Prof. Shastry led us to consider the greater context of these results in connection to Refs. \onlinecite{Kohn,Nozieres}.
This work was supported by the EPSRC through the grants EP/H049797/1 and  EP/l02669X/1.

\setlength{\bibsep}{1pt}


\begin{thebibliography}{10}
\bibitem{comment3}
Technically, the divergence comes from the integration in \Eq{SigmaU2} over the region $ k_{F2} < q < \Lambda $ 
where the Green's function can be approximated as $G(\vec k_{F2} + \vec q,\omega) \approx G(\vec q, \omega) = 1/(i \omega + q^2/2) $.
\bibitem{comment2}
This is defined as the $i$th branch of the solution of the equation $z=w \, e^w$.
\bibitem{comment4}
This implies that if we continuously vary the doping, there is an interval of electron densities where no stable homogeneous phase
exists; there would be phase coexistence (phase-separation).  The physics of such a state is  governed by long-range
interactions that are not included in the present model.  In this situation, we also cannot exclude the possibility of a charge density wave phase with broken translational symmetry.
\end{thebibliography}
\end{document}